\documentclass[dvips,11pt]{article}
\usepackage{varioref,exscale,latexsym,amsmath,amssymb}
\usepackage{graphicx}

\def\beq{\begin{equation}}

\def\eeq{\end{equation}}

\def\beqa{\begin{eqnarray}}

\def\eeqa{\end{eqnarray}}

\title{{\bf Isotropy of the early universe from CMB anisotropies}}
\author{Evan P. Donoghue $^{a,b}$ and John F. Donoghue$^a$ \\ \\
$^a$ Department of Physics\\
University of Massachusetts\\
Amherst, MA  01003\\  \\
$^b$ University of Notre Dame, \\ Notre Dame, IN 46556}
\begin{document}
\begin{titlepage}
\maketitle
\begin{abstract}
The acoustic peak in the CMB power spectrum is sensitive to causal processes and cosmological
parameters in the early universe up to the time of last scattering. We provide limits on
correlated spatial variations of the peak height and peak position and interpret these as constraints on the spatial variation of the cosmological parameters (baryon density, cold dark matter density and cosmological
constant as well as the amplitude and tilt of the original fluctuations). We utilize
recent work of Hansen, Banday and G\'{o}rski (HBG) who have studied the spatial isotropy
of the power spectrum as measured by WMAP by performing the power spectrum analysis on smaller patches of the sky.
We find that there is no statistically significant correlated asymmetry of the peak.
HBG have also provided preliminary indications of a preferred direction in the lower
angular momentum range( $\ell\sim$ 2-40) and we show how possible explanations of
this asymmetry are severely constrained by the data on the acoustic peak.
Finally we show a possible non-gaussian feature in the data, associated with a difference in the northern
and southern galactic hemispheres.
\end{abstract}
\end{titlepage}
\section{Introduction}
One of the foundations of our description of the universe is the principle of
isotropy and homogeneity - the statement that the universe is the same in all directions
and locations. The isotropy of the universe can be explained by the theory of inflation,
in which all of the observed universe descends from a single causally connected
patch of the early universe. However, inflation also contains the ingredients for generating anisotropies
on various scales. This is manifest most importantly in the temperature anisotropies of
the cosmic microwave background (CMB). The temperature anisotropies are generated by
the quantum fluctuations of a scalar field during the inflationary epoch. Other
scalar fields could also have quantum fluctuations and these could in
principle influence the distribution of matter in different portions of the universe. It
is possible that these could leave a residual ``tilt'' to the universe\cite{turner}.

Recently Hansen, Banday and G\'{o}rski (HBG) \cite{hbg, hbbg, Hansen} have explored novel
tests of isotropy by studying the WMAP\cite{WMAP}power spectrum extracted from the CMB
using patches in different directions of the sky. This is an important test because the
CMB provides the largest lever arm for the study of spatial asymmetries.
 In fact, in the lower angular
range of $\ell\sim$2-40 HBG find preliminary indications (at about the 2-3 sigma level)
for non-zero differences in the power spectrum in hemispheres oriented along galactic
co-latitude and longitude (80$^o$, 57$^o$), close to the ecliptic pole. This directional asymmetry
seems to be connected with the orientation of the quadrupole and octopole \cite{Schwarz}.
If this signal proves to be real, it could be
an important indication of a fundamental tilt in the properties of the
universe. In addition, a lack of isotropy has been seen in tests for non-gaussian behavior.
Eriksen et al \cite{Eriksen3} and Park \cite{Park} were the first to find a difference between
the northern and southern galactic hemispheres in tests for non-gaussianity, and this hemisphere
difference has subsequently been confirmed by other measures \cite{Eriksen, Eriksen2, Land2, HansenCMV}.

HBG also studied the height and the location of the first acoustic peak on different
patches of the sky, finding no obvious anisotropy within the expected cosmic variance. We
use the HBG data to study the properties of the acoustic peak in more detail. The region of the acoustic
peak is special because it is sensitive to causal processes in the early universe up to
the time of last scattering. These processes depend on the cosmological parameters and any
gradient of the parameters would yield a specific pattern between the peak height and peak position, i.e. both the height and the position would change in a correlated way. These correlations could reveal a signal of
anisotropy that is not visible when looking at the height or position
individually. We explore the HBG data for the possible presence of these correlated variations.
Interpreting the results as variations in the cosmological parameters, we provide upper bounds on their possible variation.
When we allow for an arbitrary correlated variation, we will find a weak correlation to a directions on the sky,
but we will argue that it is not statistically significant. We also use the properties of the acoustic
peak as a probe of the possible physics that could produce the low $\ell$ signal found by HBG.
The isotropy of the peak is a strong constraint on the physics that could produce the low $\ell$ signal. If one studies only the variation of a single cosmological parameter, only a difference of the spectral tilt
could explain the low-$\ell$ anomaly while being consistent with the properties of the acoustic peak.
Finally we describe an unusual behavior in the distribution of the properties
of the acoustic peak, associated with a difference between the northern and southern galactic
hemispheres. This may
possibly be related to the north-south difference which has been found in tests for non-gaussianity.

\section{Characteristic patterns of individual variations}

The acoustic peak is categorized by its location in $\ell$ and by its height. However, these are not
fully independent variables in terms of their dependence on the cosmological parameters. If one were
modify the baryon density, for example, there would be a correlated variation in both the height and location
of the peak. In this section we describe the nature of these correlations.

Consider the variation of the cosmological parameters. In all cases, we adopt as the central value
that found by the WMAP team. In addition, we define a parameter
\begin{equation}
\epsilon_i=\frac{p_i-p_i^{\rm WMAP}}{p_i^{\rm WMAP}}
\end{equation}
where $p_i$ is any of the cosmological parameters and the superscript WMAP denotes its value
measured by WMAP.

The dependence of the power spectrum on all the cosmological parameters can
be calculated using CMBFAST\cite{cmbfast}. Here we focus on the properties of the
acoustic peak. It is important to use the version of CMBFAST that does not implement the COBE normalization.
Otherwise, changes in the power at low values of $\ell$ would have an extra impact on the
amplitude of the acoustic peak by rescaling the overall power spectrum. This is
especially relevant for variations of the spectral tilt and the cosmological constant, which
appreciably modify the power at lower values of $\ell$. Taking small
variations of the cosmological parameters with modify the location of the
acoustic peak, $\ell_p$, and the height of the peak, $h_p$, from the values found by WMAP, called
$\ell_0$ and $h_0$. In each case we keep $\Omega_{\rm tot}= 1$ as we vary the parameter.
For small variations the changes will be linear in
$\epsilon_i$, so that we parameterize the results by
\begin{eqnarray}
\ell_p &=& \ell_0 + a_i\epsilon_i \nonumber \\
h_p &=& h_0 +b_i\epsilon_i
\end{eqnarray}
Thes can be combined to describe the correlation characteristic of each parameter, using
\begin{equation}
\ell_p-\ell_0 = r_i(h_p-h_0)
\end{equation}
where
\begin{equation}
r_i = \frac{a_i}{b_i}
\end{equation}
The value of the different coefficients are given in Table 1. The correlated
variations in the ($h,\ell$) plane are shown in Figure 1 for the different parameters.

   \begin{table}[h]
 {}\hspace{46pt}\begin{tabular}{c|c|c|c|c|c}
\({\rm parameter}\) & \({\rm Baryon}\) & \(CDM\) & \(\Lambda\)  & \({\rm amplitude}\)  & \({\rm tilt}\)  \\ \hline \(a\)       & \(10\) & \(0\) & \(-22\)
& \(-16\)  & \(37\)     \\ \hline
\(b\)       & \(1875\)  & \(-2381\)  & \(0\)& \(5600\)   & \(-5491\)      \\ \hline
 \end{tabular}
\caption{Parameters describing the variation of the height and location of the acoustic peak when one
changes the energy density of baryons, the energy density of cold dark matter, the value of the cosmological constant or the value of the spectral tilt. \label{a and b parameters}}
 \end{table}

\begin{figure}[h]
\begin{center}
  \begin{minipage}[t]{0.93\textwidth}
    \vspace{0pt}
    \centering
    \hspace{-40pt}\
    \hspace{-40pt}
    \includegraphics[width=0.80\textwidth,height=!]{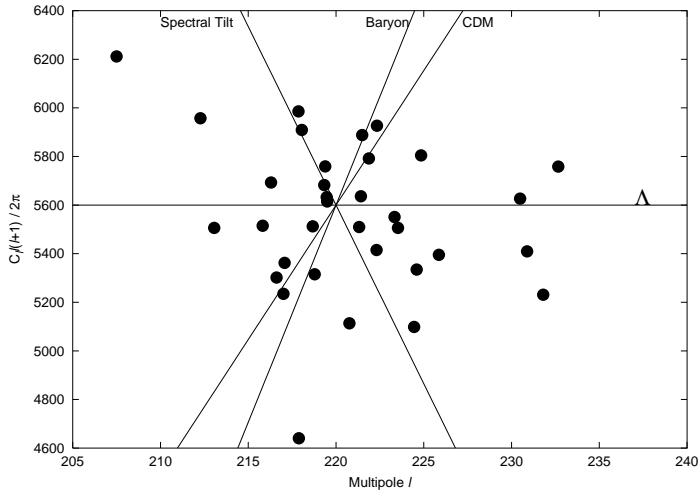}
  \end{minipage}
\end{center}
$\hspace*{0.515\textwidth}$ \caption{The lines show the correlated variation of the height and
location in $\ell$ of the acoustic peak as one varies each of the indicated quantities.
Also shown are the HBG data for the height and location of the peak for 34 patches of the sky. \label{data}
\vspace*{10pt}}
\end{figure}
\section{Measures of Isotropy}

Hansen et. al. (HBG)\cite{hbg} have studied the WMAP power spectrum on 34 patches of the sky, each with an opening angle of 19$^o$. These cover most of the sky, avoiding the galactic plane where foreground
contaminations could distort the signal. The HBG data on the location and height of the acoustic peak
is displayed in Fig 1. The correlation with position on the sky is given in Table 2. The expected spread in these values has
been estimated by HBG through the variance found in 512 simulated skies with gaussian distributed
temperature fluctuations using the WMAP power spectrum. This yields a variance in the $\ell$
direction of $\sigma_\ell = 6.5$ and in the height of $\sigma_h = 300$.
   \begin{table}[h]
 {}\hspace{30pt}\begin{tabular}{c|c|c|c}
\({\rm co-latitude}\) & \({\rm longitude}\) & \(\ell ~{\rm location}\) & \({\rm amplitude}\)   \\ \hline
\(0\)       & \(0\) & \(223.3\) & \(5551\)  \\ \hline
\(35\)       & \(0\)  & \(217.0\)  & \(5235\)     \\ \hline
\(35\)       & \(60\) & \(215.8\) & \(5515\)  \\ \hline
\(35\)       & \(120\)  & \(218.1\)  & \(5909\)     \\ \hline
\(35\)       & \(180\) & \(207.5\) & \(6211\)  \\ \hline
\(35\)       & \(240\)  & \(231.8\)  & \(5231\)     \\ \hline
\(35\)       & \(300\) & \(232.7\) & \(5758\)  \\ \hline
\(65\)       & \(0\)  & \(212.3\)  & \(5957\)     \\ \hline
\(65\)       & \(36\) & \(217.9\) & \(4640\)  \\ \hline
\(65\)       & \(72\)  & \(230.9\)  & \(5410\)     \\ \hline
\(65\)       & \(108\) & \(224.5\) & \(5099\)  \\ \hline
\(65\)       & \(144\)  & \(223.5\)  & \(5506\)     \\ \hline
\(65\)       & \(180\) & \(221.5\) & \(5888\)  \\ \hline
\(65\)       & \(216\)  & \(213.1\)  & \(5506\)     \\ \hline
\(65\)       & \(252\) & \(217.9\) & \(5986\)  \\ \hline
\(65\)       & \(288\)  & \(222.3\)  & \(5927\)     \\ \hline
\(65\)       & \(324\) & \(224.6\) & \(5335\)  \\ \hline
\(115\)       & \(0\)  & \(219.5\)  & \(5633\)     \\ \hline
\(115\)       & \(36\) & \(222.3\) & \(5415\)  \\ \hline
\(115\)       & \(72\)  & \(221.3\)  & \(5510\)     \\ \hline
\(115\)       & \(108\) & \(219.5\) & \(5615\)  \\ \hline
\(115\)       & \(144\)  & \(224.8\)  & \(5804\)     \\ \hline
\(115\)       & \(180\) & \(216.3\) & \(5693\)  \\ \hline
\(115\)       & \(216\)  & \(230.5\)  & \(5627\)     \\ \hline
\(115\)       & \(252\) & \(220.8\) & \(5113\)  \\ \hline
\(115\)       & \(288\)  & \(225.9\)  & \(5395\)     \\ \hline
\(115\)       & \(324\) & \(219.4\) & \(5759\)  \\ \hline
\(145\)       & \(0\)  & \(218.8\)  & \(5315\)     \\ \hline
\(145\)       & \(60\) & \(219.3\) & \(5683\)  \\ \hline
\(145\)       & \(120\)  & \(221.9\)  & \(5792\)     \\ \hline
\(145\)       & \(180\) & \(216.6\) & \(5302\)  \\ \hline
\(145\)       & \(240\)  & \(217.1\)  & \(5362\)     \\ \hline
\(145\)       & \(300\) & \(221.4\) & \(5636\)  \\ \hline
\(180\)       & \(0\) & \(218.7\) & \(5513\)  \\ \hline
 \end{tabular}
\caption{HBG data on the height and location of the acoustic peak on patches centered on
different locations on the sky. We thank the authors of \cite{hbg} for permission to include this data.
 \label{HBG}}
 \end{table}

A comparison of the average values of the position and height of the peak over opposite hemispheres is subject to
large fluctuations as individual patches near the borders of the hemispheres are
counted first in one hemisphere and then in the other as the orientation of the
hemisphere changes slightly in direction. Therefore we seek a measure comparing directions
that is not as sensitive to the border between the hemispheres. To this end we consider a possible dipole asymmetry. Theoretically a dipole is often the leading multipole for a spatial
asymmetry \cite{turner, Castro, Donoghue, Clifton, Boucher, Kaloper, Kurki, Langlois}.
A dipole tilt in the Universe can also be hidden beneath the overall dipole seen in the CMB temperature,
which is usually attributed to our motion with respect to the CMB, but which can also have
other components\cite{turner, Donoghue}.

Consider a reference direction given by polar angles $\theta,~\phi$
and let $\gamma$ be the angle between this reference direction and a point in the sky.
For a given parameter $p_i$, a correlated dipole variation centered on this reference
direction has the form
\begin{eqnarray}
\ell (\theta,\phi) = \ell_0 + a_i\eta \cos \gamma \nonumber \\
h(\theta,\phi) = h_0 + b_i\eta \cos \gamma
\end{eqnarray}
where $a_i,~b_i$ are the constants determined in the previous section and $\eta$
is the amplitude of a possible variation.

For each of the cosmological parameters, we allow the reference direction to vary
and search for a direction that gives the maximum possible dipole
asymmetry.\footnote{These studies were performed with MINUIT.} While in each
case a preferred direction can be found, the resulting asymmetries are
not statistically significant, as we discuss below. The amplitude and
standard deviation for each variation is shown in Table 3. For each of the
parameters, we can convert our lack of a signal into a bound. For the direction that
maximizes the asymmetry we take the 2$\sigma$ upper bound on the amplitude $\eta$.
The results are given in Table 3.

   \begin{table}[h]
 {}\hspace{-30pt}\begin{tabular}{c|c|c|c|c|c}
\({\rm parameter}\) & \({\rm Baryon}\) & \(CDM\) & \(\Lambda\)  & \({\rm amplitude}\)  & \({\rm tilt}\)  \\ \hline
\(\eta ~{\rm fit}\)       & \(0.076 \pm 0.048\) & \(0.004\pm 0.11\) & \(0.055\pm 0.035\)
& \(0.027 \pm 0.017\)  & \(0.03\pm 0.02\)     \\ \hline
\(2 \sigma ~{\rm bound}\)       & \(0.17\)  & \(0.28\)  & \(0.125\)& \(0.061\)   & \(0.057\)      \\ \hline
 \end{tabular}
\caption{Isotropy bounds when interpreted as variations of the underlying parameters. The
first row of results gives the fit value of the fractional variation of each parameter, while
the second row reports the 2 $\sigma$ upper limit on the possible variation. \label{results}}
 \end{table}

We also searched for a dipole anisotropy with an arbitrary correlation between the
height and the peak location. We do this by allowing the relative size of $a_i$ and
$b_i$ to be arbitrary. The result was an anisotropy centered on the direction
($\theta = 107^o \pm34^o, ~\phi=30^o\pm 42^o$) with a small amplitude that is nominally
1.6 standard deviations from zero and a correlation
\begin{equation}
r=\frac{a}{b} = -0.3\pm 2.0
\end{equation}
The ratio $r$ has a large uncertainty but favors a signal which is mostly in the overall
height of the amplitude. Numerically, the amplitude corresponds to $b\eta =148\pm93$,
which is a signal of less than 3\% and a nominal statistical likelihood of
1.6 standard deviations. Although this likelihood is of marginal significance,
the true significance is weaker because we have allowed an arbitrary correlation
by accepting any ratio $r$. To demonstrate this,
we performed a similar search on 100 sets of simulated data, generated randomly
with a gaussian weight and the
same standard deviations given by HBG. If we allow the signal to have an
arbitrary value or $r$ and an arbitrary direction, we find that 85\% of the simulated data
sets also have a non-zero amplitude of at least 1.6 standard deviations. We conclude that the signal
is not statistically significant.

\section{Constraints on the possible asymmetry at low $\ell$}

In the previous section, we studied the constraints imposed by the properties of the acoustic
peak. However, some of the parameters can also modify the power spectrum at lower values of $\ell$.
Indeed a study at lower values is especially interesting because HBG have identified a
potentially significant asymmetric signal for the values $\ell\sim 2-40$, with the hemisphere
oriented in the directions ($80^o,~57^o$) having about 20\% more power than in the opposite hemisphere.
Subsequent work by Hansen et.al. (HBBG)\cite{hbbg} has explored the
possibility that this signal could be due to correlated changes in the combination
of the optical depth and the initial amplitude of fluctuations, or alternatively in the combination of the
amplitude and tilt of the initial fluctuations. In this section, we explore how the
constraints from the acoustic peak impact on possible explanations of the asymmetry found by HBG.

Of the cosmological parameters that we study, the ones that influence the power spectrum at lower values of
$\ell$ are the overall amplitude, the spectral tilt and the cosmological constant. The latter is
by far the easiest to rule out as the source of the HBG asymmetry. A variation in the cosmological constant would
lead to changes at low values of $\ell$ through the late time integrated Sachs-Wolfe effect and
also lead to changes in the acoustic peak\footnote{The cosmological constant does not influence the causal
physics up to the time of last scatter, but it influences the angular scale of the acoustic peak.}.
However, in order to produce a 20\% change in the
power spectrum at low $\ell$, one would require over 100\% variation of the cosmological
constant. Our constraints given above firmly rule this out.

  The simplest variation that would explain the HBG signal is that of the
overall amplitude, requiring a 20\% larger amplitude in one hemisphere than the other. Since the whole power
spectrum is proportional to this amplitude, the same effect would show up
at the acoustic peak. This is not seen, as is clear from our overall constraint of the
previous section. Moreover, when we study the amplitude variation centered on the HBG
direction, we find an even tighter constraint, with a best fit amplitude of $\eta=0.02\pm 0.02$
and a 2$\sigma$ bound of $\eta\le 0.06$.

Finally a variation in the tilt of the CMB spectrum would also influence the
relative height of both the low and high $\ell$ regions. Here we are guided by the
final result of HBBG in the section where they consider the possibility of
correlated variations of both the tilt and the amplitude. Their central values are consistent with
the same amplitude but a tilt that is different in the two hemispheres by 8\%, a 2-3$\sigma$ effect.
We test this hypothesis by considering the properties of the acoustic peak under such variation. The spectral
tilt has the property that it influences the power spectrum more strongly at lower values of $\ell$
than it does at the acoustic peak. Our studies with CMBFAST reveal that a
$20\pm 8\%$ variation in the power at low values can be fit with a variation of the spectral
index of $0.07\pm 0.03$, and that this would be expected to
produce a variation in the height of the acoustic peak of $7\pm 3 \%$, with a corresponding change
in the peak location. We can search for such a variation, finding a difference of $3\pm 2\%$, i.e. a small but not statistically compelling signal. However, within the error bars, this observation is
compatible with the expected signal.

We also can consider the combined variation of both the amplitude and the spectral tilt, using
the information at lower values of $\ell$ as well as information at the peak. Performing
a simultaneous fit to both regions, we find consistency for
\begin{eqnarray}
n_N-n_S &=& 0.083\pm 0.041 \nonumber \\
\frac{A_N}{A_S} &=& 1.05 \pm 0.025
\end{eqnarray}
This result is consistent with that of HBBG within the error bars.

\section{A hint of non-gaussianity}

We have found no significant signal of a correlated variation of the height and location
of the acoustic peak. However the data does appear to contain a more unusual
asymmetry, which we discus in this section. The northern and southern galactic hemispheres
appear to have a different variance in the spread of the heights and locations of the peak,
with the northern hemisphere having more variation. This is visible to the eye in Fig. 2.
To probe this effect we define a variable
\begin{equation}
\delta_i= \sqrt{(\frac{\ell_i-\ell_0}{\sigma_\ell})^2+(\frac{h_i-h_0}{\sigma_h})^2}
\end{equation}
The distribution of the values of $\delta$ for the northern and southern
hemispheres is shown in Fig. 3. It is easy to see visually that there is a significant
difference between the hemispheres, with the northern hemisphere appearing to be
anomalous. We refer to this as a
non-gaussian feature since at would imply that there is
not a uniform gaussian distribution in the properties of the
peak.
\begin{figure}[h]
\begin{center}
  \begin{minipage}[t]{0.93\textwidth}
    \vspace{0pt}
    \centering
    \hspace{-80pt}
    \hspace{-40pt}
    \includegraphics[width=0.80\textwidth,height=!]{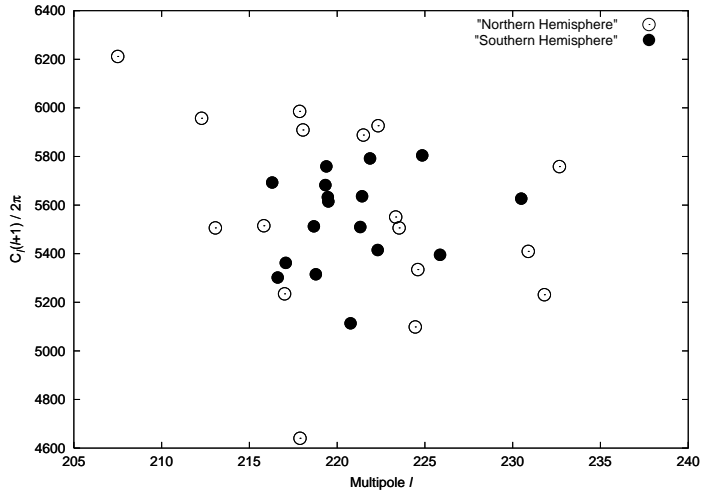}
  \end{minipage}
\end{center}
$\hspace*{0.515\textwidth} $ \caption{The data points on the acoustic peak are here separated into
those of the northern (open circles) and southern (filled circles) galactic hemispheres. One
can see visually that the data for the northern hemisphere is more
widely scattered than that of the southern hemisphere.  \label{hemispheres}
\vspace*{10pt}}
\end{figure}

We search for the direction that has the greatest non-gaussianity by identifying
the direction that maximizes the asymmetry
\begin{equation}
f(\theta,\phi) = \sum_i \delta_i \cos \gamma_i
\end{equation}
This search identifies $\theta\sim \pi$ as the direction that maximizes the difference, i.e.
there is an asymmetry between the northern and southern galactic hemispheres.
The net difference corresponds
to $f(\theta=\pi,\phi) =-7.5$.

We study the significance of this signal by simulating 256 data sets chosen randomly
with a gaussian weight and the
standard deviations given by HBG. In each case we search for the direction that maximizes the asymmetry
$f(\theta,\phi)$ and find how large the asymmetry is when centered on this direction. The simulations
reveal a median of $\bar{f}(\theta,\phi)=-3.3$ and a standard deviation of $\sigma_f = 1.3$.
Consistently, we find that only one of the 256 simulated data
sets has an asymmetry as large as is seen in the real data. We see that our signal is unlikely to be purely a
statistical fluctuation, with an exclusion at the 3$\sigma$ level.

If this signal is real physics, it appears as an extra random ingredient in the temperature
fluctuations - dominantly in the northern hemisphere. Since it involves regions
that were out of causal contact at the time of last scatter, it would
either correspond to local foreground effects in the present universe or
to physics very early in the period of inflation when these regions were causally
connected.
\begin{figure}[h]
\begin{center}
  \begin{minipage}[t]{0.93\textwidth}
    \vspace{0pt}
    \centering
   \hspace{-20pt}
  \hspace{-20pt}
    \includegraphics[width=0.8\textwidth,height=!]{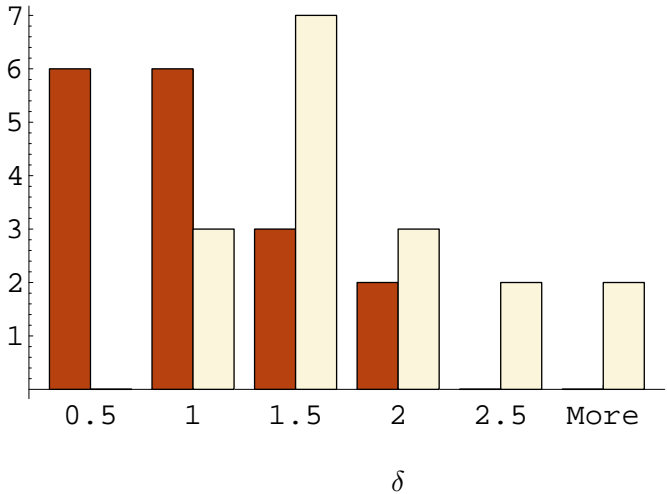}
  \end{minipage}
\end{center}
$\hspace*{0.515\textwidth} \delta$ \caption{The distribution of the parameter $\delta$ in the
northern (light) and southern hemispheres (dark). The distribution for the southern hemisphere is
consistent with the expected gaussian distribution within statistics. As described
in the text, the distribution for the northern hemisphere appears
more dispersed than expected. \label{delta}
\vspace*{10pt}}
\end{figure}
While the signal that we have identified appears to be statistically significant, it is not clear how it
is related to other measures of non-gaussian behavior. The use of other measures has produced ample
evidence for non-gaussianity in the WMAP data\cite{Eriksen3, Park, Eriksen, Eriksen2,  Land2,
HansenCMV, Land, Larson,
Vielva, Cruz, Copi, Mukherjee, McEwen, Cabella}. Moreover, as mentioned in the
introduction, for several of these signals there seems to be a
correlation with the northern and southern
hemispheres\cite{Eriksen3, Park, Eriksen, Eriksen2, Land2, HansenCMV}. For example, using a
measure of the frequency of hills, lakes or saddles in the temperature distribution, \cite{HansenCMV} found
that the northern hemisphere is non-gaussian at the 2-3 sigma level,
while no significant non-gaussianity was found in the
southern hemisphere. Our signal shares this directionality, even including the possibility that it
is the northern hemisphere that is anomalous.

\section{Conclusion}

We have studied the properties of the acoustic peak in the
CMB power spectrum and have provided measures of isotropy for the types of correlated
variations of the height and location of the peak. Within the
statistical uncertainty, no significant asymmetric signal was found.

The properties of the acoustic peak also constrain the
possible explanations of the tentative signal of an asymmetry at low $\ell$
found by HBG. If one attempts to explain this by the variation of a
single parameter from the set considered in this paper, only the
variation of the spectral tilt
is able to accommodate the constraints of the acoustic peak. This is
in agreement with the overall fit performed by HBBG\cite{hbbg}.

We have also identified a curious difference between the data in the northern and southern hemispheres,
with the northern hemisphere being more widely scattered than expected from the cosmic variance.

\section*{Acknowledgements}

We would like to thank F. Hansen for helpful comments and for providing the data listed in Table 2 and K. G\'{o}rski for useful communication. We also thank K. Dutta and A. Ross for comments and advice during the
course of this work. This work was supported in part by the U.S. National
Science Foundation.

\section*{Acknowledgement}

\end{document}